\documentstyle[12pt,epsf]{article}
\renewcommand{\bar}[1]{\overline{#1}}

\begin{document}

\begin{flushright}
USM-TH-116
\end{flushright}

\centerline{\Large \bf  Quark Fragmentation Functions in a Diquark
Model}

\centerline{\Large \bf   for  $\Lambda$ Production}

\vspace{22pt}
\centerline{\bf
Jian-Jun Yang\footnote{e-mail: jjyang@fis.utfsm.cl}$^{a,b}$}

\vspace{8pt}

{\centerline {$^{a}$Department of Physics, Nanjing Normal
University,}}

{\centerline {Nanjing 210097, China}}

{\centerline {$^{b}$Departamento de F\'\i sica, Universidad
T\'ecnica Federico Santa Mar\'\i a,}}

{\centerline {Casilla 110-V, 
Valpara\'\i so, Chile\footnote{Mailing address}}

\vspace{10pt}
\begin{center} {\large \bf Abstract}

\end{center}
Using  a simple quark-diquark model, we extract a set of
unpolarized and polarized fragmentation functions for the
$\Lambda$  based on  the available unpolarized $\Lambda$
production data in $e^+ e ^- $ annihilation. It is found that
there is a strong SU(3) flavor symmetry breaking in the
unpolarized $\Lambda$ fragmentation functions
 and that  the polarized   $u$ and $d$ quark contributions  to
 the polarized $\Lambda$ fragmentation
 are positive in medium and large $z$ region.  The 
 predictions of $\Lambda$-polarization with the
 obtained quark fragmentation functions  are compatible with  the  available
 data on the  longitudinally polarized $\Lambda$
produced in $e^+e^-$-annihilation, polarized charged lepton deep
inelastic scattering (DIS), and neutrino DIS. The spin asymmetry
for the $\Lambda$ production in $p\vec{p}$ collisions is also
predicted for a future  test in  experiments at RHIC-BNL.

\vfill
\centerline{PACS numbers: 14.20.Jn, 13.88.+e, 12.39.Ki,
13.60.Hb}

\vfill

\newpage
\section{Introduction}

Investigating the detailed structure of the nucleon is one of the
most active research directions of high energy and nuclear
physics. In order to enrich our understanding of the flavor and
spin structure of the nucleon, it is very important to apply the
same mechanisms that produce the quark structure of the nucleon to
other octet baryons and to find a new domain where the physics
invoked to explain the structure of the nucleon can be clearly
checked. However, rather little  is known about the structure of
other octet baryons. This is primarily due to their short life
time. Also one obviously cannot produce a beam of charge-neutral
hyperons such as $\Lambda$. What one can actually measure in
experiments is the  quark to $\Lambda$ fragmentation. The
$\Lambda$ hyperon is of special interest in this respect since its
decay is self-analyzing with respect to its spin direction due to
the dominant weak decay $\Lambda \to p \pi^-$ and the particularly
large asymmetry of the angular distribution of the decay proton in
the $\Lambda$ rest frame. So polarization measurements are
relatively simple to be performed and the polarized fragmentation
functions of quarks to the $\Lambda$ can be measured. Actually,
there has been some recent progress in measurements of
polarized $\Lambda$ production. The longitudinal $\Lambda$
polarization in $e^+e^-$ annihilation at the Z-pole was observed
by several collaborations at CERN [1-3]. Recently, the HERMES
Collaboration at DESY reported a result for the longitudinal spin
transfer to the $\Lambda$ in the polarized positron DIS process
~\cite{HERMES}. Also the E665 Collaboration at FNAL measured the
$\Lambda$ and $\bar{\Lambda}$ spin transfers from muon
DIS~\cite{E665}, and they observed a very different behaviour for
$\Lambda$'s and $\bar{\Lambda}$'s. Very recently, the measurement
of $\Lambda$ polarization in charged current interactions has been
done in NOMAD~\cite{NOMAD}. The high statistics investigation of
polarized $\Lambda$ production is one of the main future goals of
the HERMES Collaboration which will improve their detector for
this purpose by adding so called Lambda-wheels. In addition, we
also expect to get some knowledge of hadronization mechanism
from the experiments at the BNL Relativistic Heavy Ion Collider
(RHIC) where a polarized $pp$ collider with high luminosity and
with a center of mass (c.m.) energy $\sqrt{s} = 500~ \rm{GeV}$ is
now running~\cite{Saito}.

Much work has already been done to
relate the flavor and spin structure of the
$\Lambda$ to various fragmentation processes [8-25].
Explicit calculations have been performed for the quark
fragmentation functions  in a quark-diquark model~\cite{Nza95}.
The quark distributions inside the $\Lambda$ have also been studied 
in the MIT bag model and novel features in cases similar to the
nucleon one have been suggested~\cite{Bor99}. One of the most 
interesting  observations is related to the
polarization of  quarks inside the $\Lambda$.
In the naive quark model, the $\Lambda$ spin is exclusively
provided by the strange ($s$) quark, and the $u$ and $d$ quarks
are unpolarized. Based on novel results concerning the proton spin
structure from DIS experiments and
SU(3) symmetry in the baryon octet, it was found that the
$u$ and $d$ quarks of the $\Lambda$ should be negatively
polarized~\cite{Bur93}.
However, based on the light cone SU(6) quark diquark spectator model and
 the perturbative QCD (pQCD) counting rules analysis,
 it was found that  the $u$ and $d$ quarks should be
positively polarized at large $x$, even though their net spin
contributions to the $\Lambda$ might be zero or
negative~\cite{MSY2-3}. In order to convert the spin structure of
the $\Lambda$ into predictions for future experiments, Florian,
Stratmann and Vogelsang~\cite{Flo98b} made  a  QCD analysis of the
polarized $\Lambda$ fragmentation function within three different
scenarios. Scenario 1 corresponds to the SU(6) symmetric
non-relativistic quark model, according to which the $u$ and $d$
quark contributions to the spin of the $\Lambda$ are zero;
Scenario 2, based on a SU(3) flavor symmetry analysis and on the
first moment of $g_1$, predicts that the $u$ and $d$ quarks of the
$\Lambda$ are negatively polarized. Scenario 3 is built on the
assumption that all light quarks are positively polarized in the $\Lambda$
and contribute equally to the $\Lambda$ polarization.
 It is very interesting that the best agreement with data was
 obtained within scenario 3, {\it i.e.} for strong flavour symmetry violation.
In the above analyses, the unpolarized $\Lambda$ fragmentation
functions were usually parametrized~\cite{Flo98b} with the
assumption of SU(3) flavor symmetry  and  the polarized $\Lambda$
fragmentation functions were  proposed~\cite{Flo98b,Kot98,Bor99b}
based on simple ansatz such as $\Delta D_{q}^{\Lambda}(z)
=C_{q}(z) D_{q}^{\Lambda}(z)$ with assumed  coefficients $C_{q}(z)$,
or Monte Carlo event generators without a clear physics
motivation. In fact there is a real  need for more realistic
predictions for the  spin structure of the $\Lambda$  for future
experiments. The unpolarized $\Lambda$ fragmentation functions
have been relatively well determined by means of the  unpolarized
$\Lambda$ production in $e^+e^-$ annihilation.  
The main purpose of this work is to
extract the  $\Lambda$ fragmentation functions from the 
available experimental data with a clear physics motivation. 
We find that a
simple quark-diquark model can provide a  relationship between the
polarized and unpolarized fragmentation functions with a clear
physics picture. In addition, the model can be used to consider
the SU(3) symmetry breaking in the fragmentation functions. We
plan to propose a set of quark fragmentation functions for the $\Lambda$
based on a quark-diquark model. The unpolarized $\Lambda$ fragmentation
functions are optimized by a fit to the unpolarized cross section
of the produced $\Lambda$ in $e^+e^-$ annihilation. Then we relate
the polarized $\Lambda$ fragmentation functions to the unpolarized
ones at the initial scale within the framework of the diquark
model. Finally, we check the obtained fragmentation functions by
means of the available measurement results on the $\Lambda$ polarization.

The paper is organized as follows. In Sec.~2, we briefly describe
the quark to $\Lambda$ fragmentation functions based on a simple
quark-diqaurk model. In Sec.~3, a fit to the unpolarized cross
section of the produced $\Lambda$ in $e^+e^-$ annihilation is done
in order to optimize the unpolarized fragmentation functions for
the $\Lambda$. In Sec.~4, we calculate the $\Lambda$ polarization
in $e^+e^-$-annihilation at the $Z$ pole, the longitudinal spin
transfer to the  $\Lambda$ in polarized charged lepton DIS and the
$\Lambda$ ($\bar{\Lambda}$) polarizations in neurino
(antineutrino) DIS.  We also predict the spin asymmetry for the
$\Lambda$ production in $p\vec{p}$ collisions for a future test in
experiments. Finally, we present a discussion and summary of our
new knowledge together with our conclusions in Sec.~5.

\section{Input quark fragmentation functions in a quark-diquark model}

HERMES plans for extensive $\Lambda$ production experiments after
the HERA shut-down. It is very timely as more reliable unpolarized
and polarized $\Lambda$ fragmentation functions with a clear
physical motivation are  necessary  to serve as basis for the
analysis of the new HERMES data. We find that the diquark model
given in  Ref.~\cite{Nza95} has a clear physics picture. In this
model, the fragmentation of a quark into the $\Lambda$ is modeled
with a quark-diquark-$\Lambda$ vertex. The SU(6) spin-isospin
structure of the $\Lambda$ can be re-expressed in terms of quark
and diquark states. The SU(3) symmetry breaking can be reflected
by considering the mass difference between a scalar diquark and
vector diquark states. In addition, we will show that the polarized 
quark to $\Lambda$ fragmentation functions can be related
to the unpolarized fragmentation functions by means of some
definitive ratios given by the model.

Within the framework of the diquark  model~\cite{Nza95}, the
unpolarized valence quark to $\Lambda$ fragmentation functions can
be expressed as

\begin{equation}
D_{u_v}^\Lambda(z)=D_{d_v}^\Lambda(z)= \frac{1}{12} a_S^{(u)}(z)
+ \frac{1}{4} a_V^{(u)}(z),
\end{equation}

\begin{equation}
D_{s_v}^\Lambda(z)= \frac{1}{3} a_S^{(s)}(z),
\end{equation}
where $a_D^{(q)}(z)$  ($D=S$ or $V$) is the probability of finding
a quark $q$ splitting into $\Lambda$ with
longitudinal momentum fraction $z$ and emitting a scalar ($S$) or
axial vector ($V$) antidiquark.
The form factors for scalar
and axial vector diquark are customarily taken as the same form

\begin{equation}
\phi (k^2)=N \frac{k^2-m^2_q}{(k^2- \Lambda_0^2)^2}\label{form}
\end{equation}
with a normalization constant $N$ and a mass parameter
$\Lambda_0$. $\Lambda_0=500~\rm{MeV}$ will be adopted in our
numerical calculations.
In (\ref{form}), $m_q$ and $k$ are the mass and  the
momentum of the fragmenting quark $q$, respectively.
According to Ref.~\cite{Nza95},  $a_D^{(q)}(z)$ can be
expressed in the quark-diquark model as

\begin{equation}
a_D^{(q)}(z)=\frac{N^2 z^2 (1-z)^3}{64 \pi^2}
\frac {[2 ( M_\Lambda + m_q z)^2+
 R^2(z)]}{R^6(z)}
\end{equation}
with

\begin{equation}
R(z)=\sqrt{z m_D^2-z(1-z)\Lambda_0^2+(1-z)M_\Lambda^2},
\end{equation}
where $M_\Lambda$ and $m_D$~($D=S$ or $V$) are the mass of
the $\Lambda$ and a diquark,
respectively. In consideration of the mass difference
$M_\Lambda- M_p= 176~ \rm{MeV}$,
we choose the diquark mass $m_S=900~ \rm{MeV}$ and $m_V=1100~ \rm{MeV}$
for non-strange diquark states, $m_S=(900+176)~ \rm{MeV}$ and
$m_V=(1100+176)~ \rm{MeV}$ for diquark states  $(qs)$  with $q=u,~ d$.
The quark masses are taken as $m_u=m_d= 350~ \rm{MeV}$ and $m_s=(350+176)~
\rm{MeV}$.

Similarly, the polarized quark to $\Lambda$ fragmentation functions can
be written  as

\begin{equation}
\Delta D_{u_v}^\Lambda(z)=\Delta D_{d_v}^\Lambda(z)
= \frac{1}{12} \tilde{a}_S^{(u)}(z)
- \frac{1}{12} \tilde{a}_V^{(u)}(z),
\end{equation}

\begin{equation}
\Delta D_{s_v}^\Lambda(z)= \frac{1}{3} \tilde{a}_S^{(s)}(z),
\label{Dsv}
\end{equation}
with

\begin{equation}
\tilde{a}_D^{(q)}(z)=\frac{N^2 z^2 (1-z)^3}{64 \pi^2}
\frac {[2 ( M_\Lambda + m_q z)^2-
 R^2(z)]}{R^6(z)}
\end{equation}
for $D=S$ or $V$. What we are interested is not the
magnitude of the fragmentation functions but the flavor
and spin structure of them which are given by the diquark model.
In order to extract the flavor and spin structure information, we introduce
the flavor structure ratios

\begin{equation}
F_S^{(u/s)}(z)=\frac{a_S^{(u)}(z)}{a_S^{(s)}(z)},
\end{equation}

\begin{equation}
F_M^{(u/s)}(z)=\frac{a_V^{(u)}(z)}{a_S^{(s)}(z)},
\end{equation}
and the spin structure ratio

\begin{equation}
W_D^{(q)}(z)=\frac{\tilde{a}_D^{(q)}(z)}{a_D^{(q)}(z)},
\end{equation}
with $D=S$ or  $V$.
Then we can use the fragmentation function $D_{s_v}^\Lambda (z)$ to
express all other unpolarized and polarized fragmentation functions
as follows

\begin{equation}
D_{u_v}^\Lambda (z)=[\frac34 F_M^{(u/s)}(z) + \frac14 F_S^{(u/s)}(z) ]
D_{s_v}^\Lambda (z),
\end{equation}

\begin{equation}
\Delta D_{u_v}^\Lambda (z)=\frac 14 [ W_S^{(u)}(z)  F_S^{(u/s)}(z) -
W_V^{(u)}(z)  F_M^{(u/s)}(z) ]
D_{s_v}^\Lambda (z),
\end{equation}
and

\begin{equation}
\Delta D_{s_v}^\Lambda (z)= W_S^{(s)}(z) D_{s_v}^\Lambda (z).
\end{equation}

The quark-diquark description of the fragmentation function should
be reasonable in large $z$ region where the valence quark
contributions dominate. In small $z$ region, the sea contribution
is difficult to be included in the framework of the diquark model
itself. In order to optimize the shape of fragmentation functions,
we adopt simple functional forms

\begin{equation}
D_{s_v}^\Lambda (z) = N_s z ^{\alpha_s} (1-z) ^{\beta_s} \label{fit1}
\end{equation}
and

\begin{equation}
D_{q_s}^\Lambda (z) = D_{\bar{q}}^\Lambda (z)= \bar{N} z ^{\bar{\alpha}}
(1-z) ^{\bar{\beta}} \label{fit2}
\end{equation}
to parametrize fragmentation functions of the valence quark
$D_{s_v}^\Lambda$, sea quark $D_{q_s}^\Lambda (z)$ and antiquark
$D_{\bar{q}}^\Lambda (z)$ for $q=u, d, s$. We assume that
 $D_g^\Lambda$ and $\Delta D_g^\Lambda$, $\Delta D_{q_s}^\Lambda$, and
 $\Delta D_{\bar{q}}^\Lambda$ at the initial scale are zero and they only appear
 due to the QCD evolution. Hence, the input unpolarized and polarized
 $q \to \Lambda$ fragmentation functions  can be written as

\begin{equation}
D_{q}^\Lambda (z) = D_{q_v}^\Lambda (z) + D_{q_s}^\Lambda (z),
\end{equation}

\begin{equation}
\Delta D_{q}^\Lambda (z) = \Delta D_{q_v}^\Lambda (z).
\end{equation}

\section{A fit to unpolarized cross section of produced $\Lambda$
in $e^+e^-$ annihilation}

For a fit to the experimental data, the fragmentation functions
have to be evolved to the scale of the experiments. We used the
evolution package of Ref.~\cite{Miyama94} suitable modified for
the evolution of fragmentation functions in leading order, taking
the input scale $Q_0^2=M_\Lambda^2$ and the QCD scale parameter
$\Lambda_{QCD}= 0.3~ \rm{GeV}$.

In order to express the inclusive cross section and polarization
for the  $\Lambda$ production in $e^+e^-$ annihilation , we
introduce the following quantities

\begin{equation}
\hat{A}_q=2 \chi_{2}(v_e^2+a_e^2)v_qa_q-2 e_q \chi_1 a_q
v_e,\label{hatA}
\end{equation}

\begin{equation}
\hat{C}_q=e_q^2-2 \chi_1 v_e v_q e_q+ \chi_2 (a_e^2+v_e^2)
(a_q^2+v_q^2),\label{hatC}
\end{equation}
with
\begin{equation}
\chi_1=\frac{1}{16 \sin^2 \theta_W \cos^2 \theta_W}
\frac{s(s-M_Z^2)}{(s-M_Z^2)^2+M_Z^2\Gamma_Z^2},
\end{equation}

\begin{equation}
\chi_2=\frac{1}{256 \sin^4 \theta_W \cos^4 \theta_W}
\frac{s^2}{(s-M_Z^2)^2+M_Z^2\Gamma_Z^2},
\end{equation}
\begin{equation}
a_e=-1
\end{equation}
\begin{equation}
v_e=-1+4 \sin^2 \theta_W
\end{equation}
\begin{equation}
a_q=2 T_{3q},
\end{equation}
\begin{equation}
v_q=2 T_{3q}-4 e_q \sin^2 \theta_W,
\end{equation}
where $T_{3q}=1/2$ for $u$, while $T_{3q}=-1/2$ for $d$, $s$ quarks,
$N_c=3$ is the color number, $e_q$ is the charge of
the quark in units of the proton charge,
$s$ is the total c.m. energy squared, $\theta$ is the angle
between the outgoing quark and the incoming electron, $\theta_W$
is the Weinberg angle, and $M_Z$ and $\Gamma_Z$ are the mass and
width of $Z^0$.

\begin{figure}[htb]
\begin{center}
\leavevmode {\epsfysize=10.5cm \epsffile{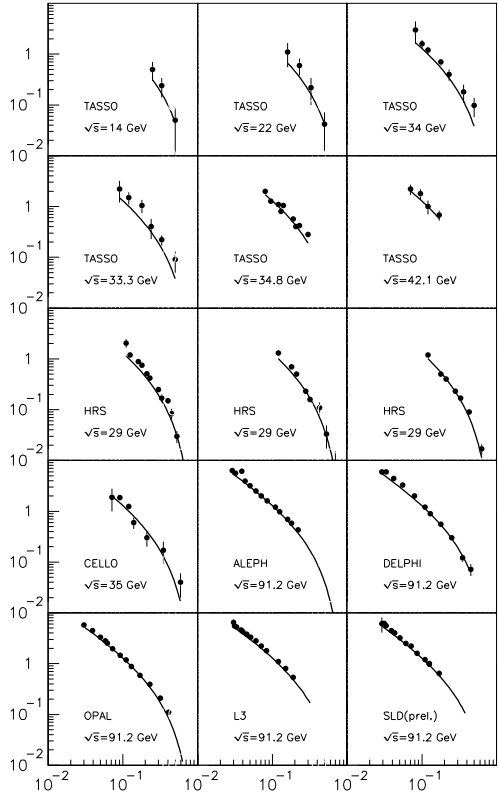}}
\end{center}
\caption[*]{\baselineskip 13pt The comparison of our results for
the $x_E$ dependence of the inclusive $\Lambda$ production cross
section $(1/\sigma_{tot})d \sigma/d x_E$ in $e^+e^-$ annihilation
and the experimental data~[26-31].
} \label{a01f1}
\end{figure}

\begin{figure}[htb]
\begin{center}
\leavevmode {\epsfysize=4.5cm \epsffile{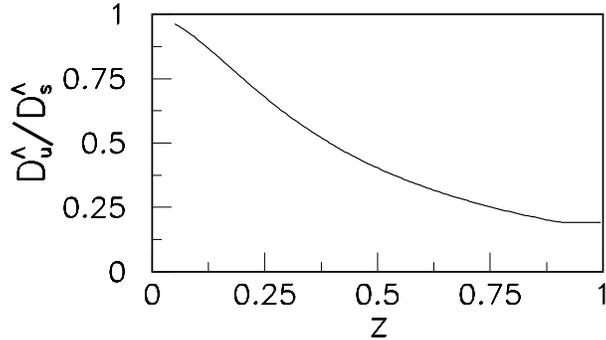}}
\end{center}
\caption[*]{\baselineskip 13pt The flavor structure of the
$\Lambda$ fragmentation functions as indicated by
$D_u^\Lambda/D_s^\Lambda$.} \label{a01f2}
\end{figure}

\begin{figure}[htb]
\begin{center}
\leavevmode {\epsfysize=4.5cm \epsffile{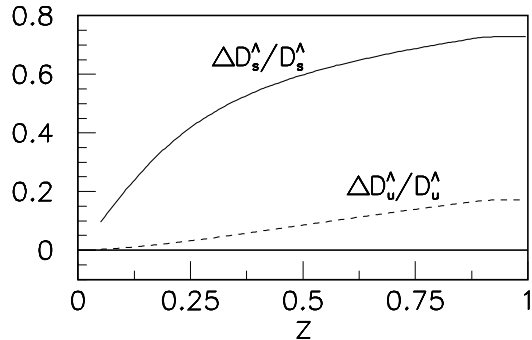}}
\end{center}
\caption[*]{\baselineskip 13pt The spin structure of the $\Lambda$
fragmentation functions as indicated by $\Delta
D_s^\Lambda/D_s^\Lambda$ (solid line) and $ \Delta
D_u^\Lambda/D_u^\Lambda$ ( dashed line).} \label{a01f3}
\end{figure}

In the quark-parton model, the differential cross section for the
semi-inclusive $\Lambda$ production process $e^+e^- \to \Lambda +
X$ can be expressed  to leading order

\begin{equation}
\frac{1}{\sigma_{tot}}\frac{d \sigma}{d x_E} =\frac{ \sum\limits_q
\hat{C}_q \left [ D_q^\Lambda (x_E,Q^2)+D_{\bar{q}}^\Lambda (x_E,Q^2)
\right ]} {\sum\limits_q \hat{C}_q} \label{crosection}
\end{equation}
where  $x_E=2 E_\Lambda/\sqrt{s}$ with  $E_\Lambda$ being
the energy of the produced $\Lambda$ in the $e^+e^-$ c.m. frame, and
$\sigma_{tot}$ is the total cross section for the process.

We perform  the leading order fit since the analysis in
Ref.~\cite{Flo98b} shows that the leading order fit can arrive at
the same fitting quality as  the next-to-leading order fit.
 By fitting the inclusive  unpolarized $\Lambda$ production data in $e^+e^-$
annihilation, the optimal  parameters in Eqs.~(\ref{fit1})-(\ref{fit2})
are obtained as  $N_s=2.5$, $\alpha_s=0.7$, $\beta_s=3.0$, $\bar{N}=0.4$,
$\bar{\alpha}=-0.51$, and $\bar{\beta}=7.4$. In Fig.~\ref{a01f1},
we show our fit results compared with the experimental
data~[26-31]. In order to show the flavor and spin structure of
the $\Lambda$ fragmentation functions, the ratios of
$D_u^\Lambda(z)/D_s^\Lambda(z)$ and $\Delta
D_s^\Lambda(z)/D_s^\Lambda(z)$ ($\Delta
D_u^\Lambda(z)/D_u^\Lambda(z)$) at $Q^2=4~ \rm{GeV}^2$ are
presented in Fig.~\ref{a01f2} and \ref{a01f3}, respectively.
   From Fig.~\ref{a01f2}, we can see that there is a strong
   flavor symmetry violation
in the unpolarized $\Lambda$ fragmentation functions, especially
at large $z$.

\section{Prediction of  spin observables in various
$\Lambda$ production processes}

There has been  available data on polarized $\Lambda$
fragmentation functions in $e^+e^-$ annihilation at the Z-pole and
also in lepton DIS. We can check our fragmentation based on these
experimental data. First, the polarized $s$ quark fragmentation
function can be checked in $e^+e^-$ annihilation at the Z-pole
since the $\Lambda$ polarization of this process is dominated by
the $s$ quark fragmentation function. Second, the spin transfer in
$\Lambda$ electroproduction is dominated by the spin transfer from
the $u$ quark to the $\Lambda$ due to the charge factor for the
$u$ quark. Morever, due to isospin symmetry the $u$ and $d$ quark
spin transfers to the $\Lambda$ are expected to be equal. We can
check the $u$ and $d$ quark fragmentation functions by means of
the $\Lambda$ production in the polarized charged lepton DIS
process. Third, the very recent NOMAD data on the $\Lambda$
polarization in the neutrino DIS process, which  has small errors,
can help us to draw a clear distinguish between different
predictions.
 Finally, the spin transfer in the
reaction $p \vec{p} \to \vec{\Lambda} X$ at RHIC BNL should be
also a good tool to discriminate between various sets of polarized
fragmentation functions compatible with the LEP data.
In order to check the obtained quark
fragmentation  functions, we apply them to predict the  spin
observables in the above various $\Lambda$ production processes.

\subsection{$\Lambda$ polarization in $e^+e^-$-annihilation}

One interesting feature of quark-antiquark ($q \bar q$) production
in $e^+e^-$-annihilation near the $Z$-pole is that the produced
quarks (antiquarks) are polarized due to the interference between
the vector and axial vector couplings in the standard model of
electroweak interactions, even though the initial $e^+$ and $e^-$
beams are unpolarized. There have been measurements of the
$\Lambda$-polarization near the $Z$-pole~[1-3].
Theoretically,  the $\Lambda$-polarization can be expressed as

\begin{equation}
P_{\Lambda}=-\frac{\sum\limits_{q} \hat{A}_q [\Delta D_q^\Lambda
(z)-\Delta D_{\bar q}^\Lambda (z)]}{\sum\limits_{q} \hat{C}_q
[D_q^\Lambda (z)+ D_{\bar q}^\Lambda (z)]}. \label{PL2}
\end{equation}
where $\hat{A}_q$ and $\hat{C}_q$ ($q=u, d$ and $s$) are given in
(\ref{hatA}) and (\ref{hatC}), respectively. Our theoretical
prediction for the $\Lambda$ polarization at the $Z$-pole is shown in
Fig.~\ref{a01f4} together with the experimental data. From
Fig.~\ref{a01f4}, we can see that the $\Lambda$ polarization,
which is mainly shaped by the polarized $s$ quark to $\Lambda$
fragmentation (see Fig.~\ref{a01f3}), is consistent with the
experimental data.

\begin{figure}[htb]
\begin{center}
\leavevmode {\epsfysize=5.5cm \epsffile{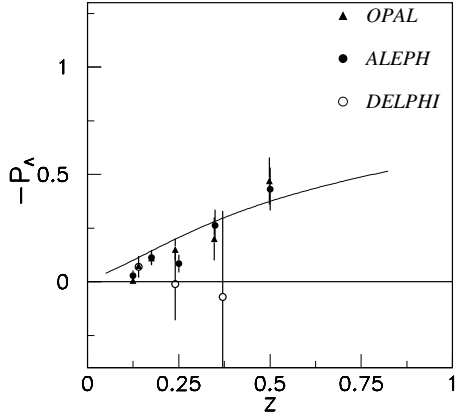}}
\end{center}
\caption[*]{\baselineskip 13pt The comparison of the experimental
data~[1-3]
for the longitudinal $\Lambda$-polarization $P_{\Lambda}$ in
$e^+e^-$-annihilation  at the $Z$-pole with our theoretical
prediction.} \label{a01f4}
\end{figure}

\subsection{Spin transfer to $\Lambda$ in polarized charged lepton DIS}

For a longitudinally polarized charged
lepton beam and an unpolarized target, the $\Lambda$ polarization
along its own momentum axis is given in the quark parton model
by~\cite{Jaf96}
\begin{equation}
P_{\Lambda}(x,y,z) = P_B D(y)A^{\Lambda}(x,z)~,
\label{PL}
\end{equation}
where $P_B$ is the polarization of the charged lepton beam, which
is of the order of 0.7 or so~\cite{HERMES,E665}. $D(y)$, whose
explicit expression is
\begin{equation}
D(y)=\frac{1-(1-y)^2}{1+(1-y)^2},
\end{equation}
is commonly referred to as the longitudinal depolarization factor
of the virtual photon with respect to the parent lepton, and
\begin{equation}
A^{\Lambda}(x,z)= \frac{\sum\limits_{q} e_q^2 [q^N(x,Q^2) \Delta
D_q^\Lambda(z,Q^2) + ( q \rightarrow \bar q)]}
{\sum\limits_{q} e_q^2 [q^N (x,Q^2)
D^\Lambda_q(z,Q^2) + ( q \rightarrow \bar q)]}~,
\label{DL}
\end{equation}
is the longitudinal spin transfer to the $\Lambda$. Here $y=\nu/E$
is the fraction of the incident lepton's energy that is
transferred to the hadronic system by the virtual photon.
 In Eq.~(\ref{DL}), $q^N(x,Q^2)$,  the quark distribution of the proton,
is adopted as  the CTEQ5 set 1 parametrization form~\cite{CTEQ5}
 in our numerical calculations.
As shown in  Fig.~\ref{a01f5}, our prediction is compatible with the
available experimental data in medium $z$ region, which suggests
that the $u$ and $d$ quark to the $\Lambda$ fragmentation
functions  are likely positive polarized in medium and large $z$ region. 
However, the experimental data in small $z$ region, whose precision is  still
poor, can not be  explained at the moment.

\begin{figure}
\begin{center}
\leavevmode {\epsfysize=4.5cm \epsffile{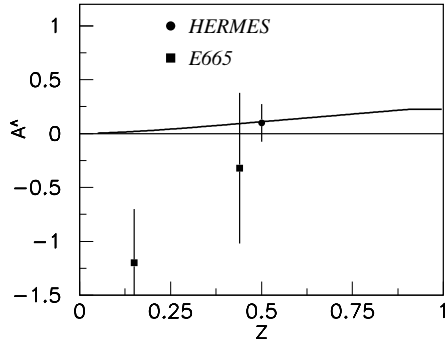}}
\end{center}
\caption[*]{\baselineskip 13pt The $z$-dependence of the $\Lambda$
spin transfer in electron or positron (muon) DIS.  Note that for
HERMES data the $\Lambda$ polarization is measured along the
virtual-photon momentum, whereas for E665 it is measured along the
virtual-photon spin. The averaged value of the Bjorken variable is
chosen as $x=0.1$ (corresponding to the HERMES averaged value) and
the calculated result is not sensitive to a different choice of
$x$ in the small $x$ region (for example, $x=0.005$ corresponding
to the E665 averaged value). $Q^2=4~\rm{GeV}^2$ is used and the
$Q^2$ dependence of the result is very weak.}\label{a01f5}
\end{figure}

\subsection{$\Lambda$ polarization in neutrino/antineutrino
DIS}

The scattering of a neutrino beam on a  hadronic target provides a
source of polarized quarks with specific flavor structure, and
this particular property makes the neutrino (antineutrino) process
an ideal laboratory to study the flavor-dependence of quark to
hadron fragmentation functions, especially in the polarized case.
We find that the $\Lambda$ polarization in the neutrino
(anti-neutrino) DIS process can also be used to check the $u$ and
$d$  quark contributions to the polarized $\Lambda$ fragmentation.

\begin{figure}
\begin{center}
\leavevmode {\epsfysize=7cm \epsffile{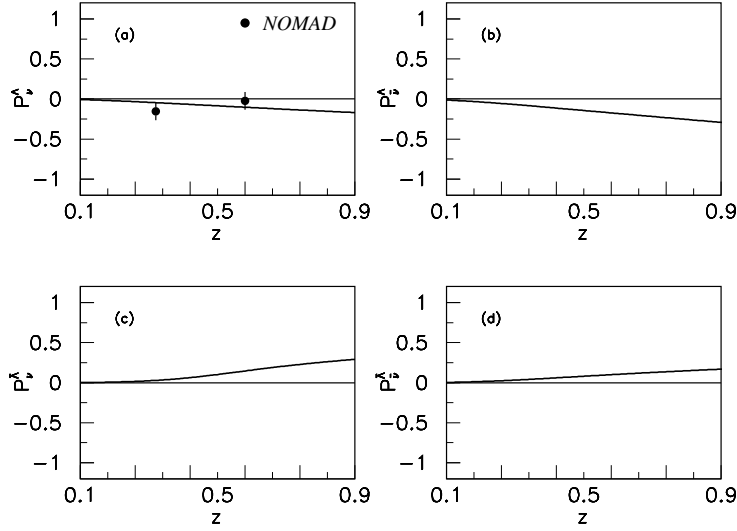}}
\end{center}
\caption[*]{\baselineskip 13pt The predictions of $z$-dependence
for the $\Lambda$ and $\bar{\Lambda}$ polarizations in the
neutrino (antineutrino) DIS process. We adopt the CTEQ5 set 1
quark distributions~\cite{CTEQ5} for the target proton at
$Q^2=4$~GeV$^2$ with the Bjorken variable $x$ integrated over
$0.02 \to 0.4$ and $y$ integrated over $0 \to 1$. }\label{a01f6}
\end{figure}

The longitudinal polarizations of the $\Lambda$  in its momentum
direction, for the  $\Lambda$ in the current fragmentation region
can be expressed as,

\begin{equation}
P_\nu^\Lambda (x,y,z)=-\frac{[d(x)+\varpi s(x)] \Delta D
_u^\Lambda (z) -( 1-y) ^2 \bar{u} (x) [\Delta D _{\bar{d}}^\Lambda
(z)+\varpi \Delta D_{\bar{s}}^\Lambda (z)]} {[d(x)+\varpi s(x)]
D_u ^\Lambda (z) + (1-y)^2 \bar{u} (x) [D _{\bar{d}}^\Lambda
(z)+\varpi D_{\bar{s}}^\Lambda (z)]}~,\label{neu1}
\end{equation}

\begin{equation}
P_{\bar{\nu}}^\Lambda (x,y,z)=-\frac{( 1-y) ^2 u (x) [\Delta D
_d^\Lambda (z)+\varpi \Delta D _s^\Lambda (z)]-[\bar{d}(x)+\varpi
\bar{s}(x)] \Delta D _{\bar{u}}^\Lambda (z)}{(1-y)^2 u (x) [D
_d^\Lambda (z)+\varpi D _s^\Lambda (z)]+[\bar{d}(x)+\varpi
\bar{s}(x)] D_{\bar{u}} ^\Lambda (z)}~,\label{neu2}
\end{equation}
where the terms with the factor $\varpi=\sin^2 \theta_c/\cos^2
\theta_c$ ($\theta_c$ is the Cabibbo angle) represent Cabibbo
suppressed contributions. There are similar formulae as
(\ref{neu1})-(\ref{neu2}) for the $\bar{\Lambda}$ whose  quark
fragmentation functions can be obtained according to the  matter
and antimatter symmetry, {\it{i.e.}} $D_{q,\bar{q}}^{\Lambda} (z)
=D _{\bar{q},q}^{\bar{\Lambda}}(z)$  and similarly for
 $\Delta D_{q,\bar{q}}^{\Lambda}(z)$.

The NOMAD data~\cite{NOMAD} on the $\Lambda$ polarization in the
neutrino DIS process, which has much smaller errors than the data
on the  longitudinal spin transfer to the $\Lambda$  in polarized
charged lepton DIS, allows to have a further check  on the
$\Lambda$ fragmentation functions. In Fig.~\ref{a01f6}, we present
our predictions of $z$-dependence for the $\Lambda$ and
$\bar{\Lambda}$ polarizations in the neutrino (antineutrino) DIS
process and find that our prediction of the $\Lambda$
polarization in neutrino DIS is consistent with the  very recent NOMAD
data~\cite{NOMAD}.

\subsection{Spin asymmetry for $\Lambda$ production in $p\vec{p}$ collisions}

In order to enrich the spin knowledge  of the nucleon and investigate
the hadronization mechanism, many spin physics programs
will be undertaken at  RHIC-BNL~\cite{Saito,Bunce}.
Theoretically,  it has been recently noticed that the $\Lambda$
polarization in polarized proton-proton
 collisions is also  very sensitive to the property of $\Lambda$
 fragmentation~\cite{Flo98, Bor99b, MSSY9}. In leading order perturbative
 QCD, the rapidity differential polarized cross section can be
 schematically written in a factorized form as

 \begin{equation}
 \frac{ d\Delta \sigma^{p\vec{p} \to \vec{\Lambda} X} }{d y}
 = \int \limits_{p_T^{min}} d p_T \sum \limits_{f^A_a f^B_b
 \to c X^\prime} \int dx_a dx_b dz_c f^A_a (x_a, \mu^2)
 \Delta f_b^B (x_b,\mu^2) \Delta D_c^\Lambda (z_c,\mu^2)
 \frac{d \Delta \hat{\sigma}}{d y} \label{pp1}
 \end{equation}
where, $f_a^A (x_a,\mu^2)$ and $\Delta f^B_b (x_b,\mu^2)$ are
unpolarized and polarized distribution functions of partons $a$
and $b$ in protons $A$ and $B$, respectively, at the scale
$\mu^2=p_T^2$. $\Delta D_c^\Lambda( z_c, \mu ^2)$ is the polarized
fragmentation function of parton $c$ into $\Lambda$ with the
momentum fraction $z_c$ of parton $c$. $y$ and $p_T$ are the
rapidity and transverse momentum of the produced $\Lambda$.
$\frac{d \Delta \hat{\sigma}}{d y}$ is the rapidity differential
polarization cross section for the sub-process $a+b \to c + d$.
The directly observable quantity is the spin asymmetry defined by
\begin{equation}
A^\Lambda_{pp}= \frac{d\Delta \sigma^{p\vec{p} \to \vec{\Lambda}
X}/dy} {d \sigma^{pp \to \Lambda X}/dy }
\end{equation}
where the unpolarized cross section $d \sigma^{pp \to \Lambda
X}/dy$ is obtained by an expression similar to the one in
(\ref{pp1}), with all $\Delta$'s removed.

\begin{figure}
\begin{center}
\leavevmode {\epsfysize=5.5cm \epsffile{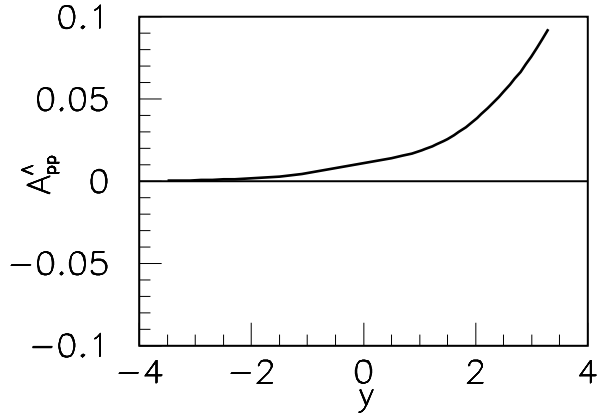}}
\end{center}
\caption[*]{\baselineskip 13pt The predictions of the
spin asymmetry as a function of rapidity
for the $\Lambda$ production in $p\vec{p}$ collisions.
 }\label{a01f7}
\end{figure}

The  result of the spin asymmetry in Fig.~\ref{a01f7} is obtained by adopting
the LO set of unpolarized parton distributions of Ref.~\cite{GRV95} and polarized parton distributions of LO
GRSV "standard" scenario~\cite{GRSV96}.
The total c.m. energy $\sqrt{s}=500~ {\rm{GeV}}$ and the lower cutoff of the
transverse momentum $p_T^{min}= 13~ {\rm{GeV}} $ are taken. By comparing the result in
Fig.~\ref{a01f7} and the spin structure of the fragmentation
function shown in Fig.~\ref{a01f3}, we can find that the spin
asymmetry for the $\Lambda$ production is sensitive to
$\Delta D_u^\Lambda/D_u^\Lambda$. Therefore, the rapidity dependence of
the spin asymmetry for the $\Lambda$ production in $p\vec{p}$ collisions
can provide a good tool to discriminate between various sets of
polarized  $u$ and $d$ quark fragmentation functions compatible
with the LEP data.

\section{Discussion and summary}

The available experimental data on the $\Lambda$ polarization 
supports our predition that the $u$ and $d$ quark contributions
to the polarized $\Lambda$ fragmentation are positive in medium and 
large $z$ region.
However, this can not be regarded as in contradiction with
the result in Ref.~\cite{Bur93}.
First, in Ref.~\cite{Bur93} the $u$ and $d$ quark contributions to
the spin content  of the $\Lambda$ were discussed. And here, we talk
about the polarized $u$ and $d$ quark
fragmentation to the $\Lambda$. There has  not been yet
a definitive relationship  between the fragmentation function
and the corresponding quark distribution function. Second, the result
in Ref.~\cite{Bur93} was obtained with the assumption of
SU(3) symmetry in the baryon octet. Recently, it has been
noticed that the effect of the SU(3) symmetry breaking in
 hyperon semileptonic decay (HSD)
 should be  significant~\cite{KimPRD, Manohar98}.
 The effect has been estimated  by a chiral quark
 soliton model~\cite{KimPRD} and the large
 $N_c$ QCD~\cite{Manohar98}. The
 consistent results  were obtained  separately by different approaches.
 The effect of the SU(3) symmetry breaking in HSD on the spin content of
 the $\Lambda$ has been  considered in the chiral quark soliton
 model~\cite{KimAPPB}. It was
 found that the integrated polarized quark
 densities for the $\Lambda$ hyperon should be
 $\Delta U=\Delta D= -0.03 \pm 0.14$ and $\Delta S=0.74  \pm 0.17$
  in the chiral limit case.
 When the strange quark mass correction is added,
 $\Delta U=\Delta D= -0.02 \pm 0.17$ and $\Delta S=1.21  \pm 0.54$.
Therefore, there are various possibilities for the
polarization of the $u$ and $d$ quarks in the $\Lambda$,
{\it{i.e.}} the $u$ and $d$ quark  contributions to the spin of the
$\Lambda$ might be zero, negative or positive.
With a statistical model~\cite{Bhalerao00}, we find that the
available experimental data on the $\Lambda$ polarization also suggests
that the $u$ and $d$ quarks are positively polarized. Therefore, the
situation in which  the $u$ and $d$ quarks are positively polarized in
the $\Lambda$ seems to be model independent.

We would like to mention that  our present knowledge on the
$\Lambda$ fragmentation functions is still poor and there are many
unknowns to be explored before we can arrive at some definitive 
conclusion on the quark spin structure of the $\Lambda$. Despite
the experimental uncertainties, it seems that the experimental
data on the longitudinal spin transfer to the $\Lambda$ in the
polarized charged lepton DIS process shows a strong dependence on
$z$, especially in low $z$ region. At  the moment, it seems to be
difficult to understand such a rather strong $z$ dependence in low
$z$ region with the available models~[14,16,20,22], although the
models can provide predictions which are compatible with the data in
medium $z$ range. Further studies in order to improve our
predictions in small $z$ region are in progress. On
the other hand, experimentally, the high statistics measurements
on the $\Lambda$ polarization have  been regarded as a strongly 
emphasized project in the COMPASS experiment~[40]. Many efforts, both
theoretically and experimentally, are being made in order to
reduce the uncertainties in the spin structure of the $\Lambda$
since the subject is crucial important for enriching the knowledge
of hadron structure and hadronization mechanism.

In summary,  we proposed a set of $q \to \Lambda$  fragmentation
functions whose flavor and spin structure are provided  by the
quark-diquark model. Using the diquark model, we related the
polarized fragmentation functions to the unpolarized ones at the
initial scale.  We optimized the fragmentation functions by
fitting  the unpolarized cross section for the $\Lambda$ produced
in $e^+e^-$ annihilation. It is found that the SU(3) symmetry
breaking in the unpolarized $\Lambda$ fragmentation functions is
significant, especially in large $z$ region. As a check of 
the obtained fragmentation functions, our predictions
for $\Lambda$ polarization in $e^+e^-$-annihilation and the
neutrino  DIS  are consistent with  the experimental
data. The prediction on the spin transfer to the $\Lambda$ in  
polarized charged lepton DIS  is compatible with the data in
medium $z$ region.  For a future test in experiments at BNL RHIC,
we predicted the spin asymmetry for the $\Lambda$ production in
$p\vec{p}$ collisions. It is found that our prediction for the
spin asymmetry is very similar to that of scenario 3 in
Ref.~\cite{Flo98b}.

{\bf Acknowledgments: } I am grateful to P. J. Mulders and J.
Rodrigues for their kindness correspondence about the
quark-diquark model. This work is stimulated by my other
cooperative work with Bo-Qiang Ma, Ivan Schmidt, and Jacques
Soffer. I also would like to express my great thanks to them for
their encouragement and valuable comments. In
addition, this work is partially supported by National Natural
Science Foundation of China under Grant Number 19875024 and by
Fondecyt (Chile) project 3990048.

\end{document}